\documentclass[11pt]{article}
\usepackage{titlesec}
\usepackage{hyperref}

\titleclass{\subsubsubsection}{straight}[\subsection]

\usepackage[utf8]{inputenc}
\usepackage{amsmath,amsfonts,amssymb,stackengine,graphicx}
\title{pre-inflationary non gaussianities}
\author{sagnotti}
\date{June 2020}

\usepackage[utf8]{inputenc}
\newif\iffigs\figstrue

\usepackage[title]{appendix}
\usepackage{epsfig,latexsym}
\usepackage{hyperref}
\usepackage{amsmath}
\usepackage{verbatim}
\usepackage{color}
\usepackage{mathrsfs}
\usepackage{slashed}
\usepackage{amssymb}

\DeclareMathAlphabet{\mathpzc}{OT1}{pzc}{m}{it}

 \csname
@addtoreset\endcsname{equation}{section}

\def\gz0{\gamma^{0}}

\def\beq{\begin{equation}}
\def\eeq{\end{equation}}
\def\bea{\begin{eqnarray}}
\def\eea{\end{eqnarray}}
\def\ba{\begin{array}}
\def\ea{\end{array}}
\def\bec{\begin{center}}
\def\ec{\end{center}}
\def\ba{\begin{align}}
\def\ena{\end{align}}

\def\12{\frac{1}{2}}

\newcounter{subsubsubsection}[subsubsection]
\renewcommand\thesubsubsubsection{\thesubsubsection.\arabic{subsubsubsection}}
\titleformat{\subsubsubsection}
  {\normalfont\normalsize\bfseries}{\thesubsubsubsection}{1em}{}
\titlespacing*{\subsubsubsection}
{0pt}{3.25ex plus 1ex minus .2ex}{1.5ex plus .2ex}

\makeatletter
\renewcommand\paragraph{\@startsection{paragraph}{5}{\z@}%
  {3.25ex \@plus1ex \@minus.2ex}%
  {-1em}%
  {\normalfont\normalsize\bfseries}}
\renewcommand\subparagraph{\@startsection{subparagraph}{6}{\parindent}%
  {3.25ex \@plus1ex \@minus .2ex}%
  {-1em}%
  {\normalfont\normalsize\bfseries}}
\def\toclevel@subsubsubsection{4}
\def\toclevel@paragraph{5}
\def\toclevel@paragraph{6}
\def\l@subsubsubsection{\@dottedtocline{4}{7em}{4em}}
\def\l@paragraph{\@dottedtocline{5}{10em}{5em}}
\def\l@subparagraph{\@dottedtocline{6}{14em}{6em}}
\makeatother

\begin{document}

\begin{flushright}
{\today}
\end{flushright}

\vspace{10pt}

\begin{center}

%%%%%%%%%%%%%%%%%%%%%%%%%%%%%%%%%%%%%%%%%%%%%%%%%%%%%%%%%%%%%%%%%%%%

{\Large\sc On Pre--Inflationary non Gaussianities}\\

%%%%%%%%%%%%%%%%%%%%%%%%%%%%%%%%%%%%%%%%%%%%%%%%%%%%%%%%%%%%%%%%%%%%

\vspace{25pt}
{\sc M.~Meo  \ and \ A.~Sagnotti\\[15pt]

{${}^c$\sl\small
Scuola Normale Superiore and INFN\\
Piazza dei Cavalieri, 7\\ 56126 Pisa \ ITALY \\
e-mail: {\small \it mario.meo@sns.it, sagnotti@sns.it}}\vspace{10pt}
}

%%%%%%%%%%%%%%%%%%%%%%%%%%%%%%%%%%%%%%%%%%%%%%%
\vspace{40pt} {\sc\large Abstract}\end{center}
%%%%%%%%%%%%%%%%%%%%%%%%%%%%%%%%%%%%%%%%%%%%%%%
\noindent
We explore the three--point amplitude of curvature perturbations in scenarios suggested by high--scale supersymmetry breaking in String Theory, where the inflaton is forced to climb a steep exponential potential. We can do it at the price of some simplifications, and more importantly with some assumptions on the softening effects of String Theory. These suggest a scenario proposed long ago by Gasperini and Veneziano, where the initial singularity is replaced by a bounce, and the resulting analysis rests on a scale $\Delta$ that leaves some signs in the angular power spectrum of the CMB. The amplitude comprises two types of contribution: the first oscillates around the original result of Maldacena and gives no further prospects to detect a non--Gaussian signal, but the second, which is subtly tied to the turning point at the end of the climbing phase,  within the window $62<N<66$ for the inflationary $e$-folds could be compatible with {\it Planck} constraints and potentially observable. The amplitudes involving the tensor modes contain only the first type of contribution.

\setcounter{page}{1}

\pagebreak

\newpage
 \newpage
\baselineskip=20pt
%%%%%%%%%%%%%%%%%%%%%%%%%%%%%%%%%%
\section[Introduction]{\sc  Introduction}\label{sec:intro}
%%%%%%%%%%%%%%%%%%%%%%%%%%%%%%%%%%%%%

The three ten--dimensional non--supersymmetric strings free of tachyons~\cite{so1616,susy95,usp32} are a good laboratory to explore the effects of supersymmetry breaking at high scales in String Theory~\cite{stringtheory}. They all share a common feature, the emergence of an exponential ``tadpole potential'' 
\beq
    V(\varphi)\ = \ \frac{T}{2 \kappa^2}\ e^{2 \,\gamma\,\varphi} \label{Vgamma_int}
\eeq
that forbids a flat vacuum,
where $\varphi$ denotes a conveniently rescaled dilaton. This occurs at the torus level for the heterotic model of~\cite{so1616}, for which $\gamma=\frac{5}{3}$, and at the (projective-)disk level for the two orientifold models~\cite{orientifolds} of \cite{susy95,usp32}, for which $\gamma=1$. A striking change of behavior~\cite{climbing} occurs in the resulting spatially flat cosmologies~\footnote{These solutions were previously found in~\cite{dm} for the potentials emerging from the ten--dimensional models, and then in~\cite{russo}, but the peculiar transition for increasing values of $\gamma$ was first noted in~\cite{climbing}.} as the parameter $\gamma$ in eq.~\eqref{Vgamma_int} is increased and reaches the value corresponding to the orientifold models, which thus concerns all three cases. The scalar field is then forced to emerge from the initial singularity while \emph{climbing up} the exponential potential, so that the resulting dynamics can be confined to the weak--coupling regime of String Theory and includes a turning point. If the tadpole potential is supplemented by a milder contribution, for example the Starobinsky potential of~\cite{starobinsky}, this setup can induce the onset of inflationary slow--roll~\cite{inflation}. 

A pre--inflationary period of fast-roll introduces a low--frequency cut in the primordial power spectrum of curvature perturbations. If this region were accessible to us, the cut would depress the first few multipoles in the CMB angular power spectrum~\cite{dkps}, which resonates with the lack of power present in them. The cut can be modeled, as in~\cite{dkps}, by deforming the Chibisov--Mukhanov primordial power spectrum~\cite{cm} into~\footnote{Mechanisms to suppress the first few CMB multipoles were previously considered in other contexts, including in~\cite{peloso}. }
\beq
P(k) \ = \ A \ \frac{\left(\frac{k}{k^\star}\right)^3}{\left[ \left(\frac{k}{k^\star}\right)^2 \ + \ \left(\frac{\Delta}{k^\star}\right)^2 \right]^{2 - \frac{n_s}{2}}} \ , \label{power_delta_int}
\eeq
where $k^\star$ is a pivot scale,
at the price of introducing the new scale $\Delta$ into the problem. Some evidence for the scale $\Delta$ was actually detected in CMB data, and reaches the 3$\sigma$ level if the masked region around the Galactic plane is widened, thus working within a sky fraction of about 39\%~\cite{gkmns}. The resulting value is of the order of the cosmic Horizon and translates into an energy scale of about $10^{12} \,e^{N-60}$ GeV at the beginning of an inflationary period of $N$ $e$-folds. However, other features of the power spectrum that are typical of the climbing scenario appear beyond reach. For example, the detection in CMB data of a pre-inflationary peak reflecting the presence of a turning point was attempted in~\cite{gkmns} using a simplified model, but did not yield significant results.

The three--point amplitude contains more information
on the early stages of the Universe, since it rests on the Green functions before and during the inflationary phase. From a theoretical point of view, the analysis of new cases has some interest of its own, since it can perhaps add to the many lessons that are being drawn from cosmological amplitudes~\cite{review_pst}, and the setting we are exploring does contain an interesting novelty. This is the presence of a turning point for the scalar where the slow--roll parameter $\epsilon$ vanishes, making the three--point amplitude based on the Mukhanov--Sasaki variable apparently singular. Or, if you will, making curvature and scalar perturbations somehow inequivalent. If one insists on working with curvature perturbations, which are closer in spirit to the actual observations, the singularity is resolved by the Schwinger--Keldysh contour, and finite contributions emerge from the neighborhood of the turning point, with peculiar and potentially interesting features. 

The initial singularity, which is also present in the cosmologies under scrutiny, is another story. All we shall be able to do concretely is appeal to the softening effects of String Theory, arguing for its eventual disappearance in a full treatment. Still, the way to do this will not be completely arbitrary, and we shall be drawn to a scenario that resonates with the ``Pre--Big Bang Cosmology" proposed long ago, in String Theory, by Gasperini and Veneziano~\cite{gasp_ven}.

With this important proviso, this paper tries to add something on the important issue of non--Gaussianities, but for brevity, we only discuss in detail the equilateral configuration, which yields the strongest signal. As we shall see, the scale $\Delta$ introduces oscillations in the dimensionless observable $f_{NL}$ originally computed by Maldacena in~\cite{maldacena} (for reviews, see~\cite{reviews}). Moreover, the contribution from the turning point identifies a narrow window in the number of inflationary $e$-folds around $N=63$, where the complete result is sizable and yet lies within limits that emerged from the {\it Planck} collaboration~\cite{Planck,forecasts}. However, as we have anticipated, all this comes at the price of appealing to the role of String Theory to regulate the pre--inflationary dynamics, in a way that we shall try to motivate and appears somewhat natural to us, but cannot be fully justified.

%%%%%%%%%%%%%%%%%%%%%%%%%%%%%%%%%%
\section[Exponential Potentials and Climbing]{\sc  Exponential Potentials and Climbing}\label{sec:climbing}
%%%%%%%%%%%%%%%%%%%%%%%%%%%%%%%%%%%%%

Consider the action principle for Einstein gravity minimally coupled to a real scalar field with a potential $V$,
\beq
    { \cal S} \ = \ \int d^4 x \sqrt{-g}\left[\frac{R}{2\,\kappa_D^2}\ - \ \frac{1}{2}\,g^{\mu \nu }\partial_{\mu}\phi\,\partial_{\nu}\phi\ -\ V(\phi)\right] \ . \label{2der_act}
\eeq
While the climbing mechanism emerged in ten--dimensional strings, here we shall work directly with its possible manifestation in four dimensions, referring mostly to the two ten--dimensional orientifold models where the phenomenon sets it. This step affords a partial justification since, if the internal volume is somehow stabilized, a ten--dimensional tadpole potential that is ``critical'' in the sense that we shall briefly recall translates into a four--dimensional one that is also critical, as explained in~\cite{fss}.

In four--dimensional spatially flat cosmologies of the form
\beq
    ds^2 \ =\  - \ e^{2B(\xi)}d\xi^2 \ + \ e^{2A(\xi)}d\vec{x} \cdot d \vec{x}\ ,
\eeq
which encode the two key properties of homogeneity and isotropy, with potentials that never vanish one can make the gauge choice~\cite{dm}
\beq
    V(\phi)\,e^{2B}\ = \ \frac{M^2}{2\,\kappa^2} \ . \label{gauge_choice}
\eeq
Combining it with the redefinitions
\beq
    \tau \ = \ \xi \,M \sqrt{\frac{3}{2}} \ , \qquad  \varphi \ = \  \kappa \,\phi\, \sqrt{\frac{3}{2}} \ , \qquad A \ = \ \frac{1}{3}\,a 
\eeq
leads to a neat form for the background equations in an expanding Universe,
\bea
    && \frac{da}{d\tau} \ =\ \sqrt{1\ +\ \left(\frac{d\varphi}{d\tau}\right)^2} \ , \nonumber \\
    &&\frac{d^2\varphi}{d\tau^2}\ +\ \frac{d\varphi}{d\tau}\sqrt{1\ +\ \left(\frac{d\varphi}{d\tau}\right)^2}\ +\ \frac{1}{2V}\frac{\partial V}{\partial \varphi}\left[1\ +\ \left(\frac{d\varphi}{d\tau}\right)^2\right] \ = \ 0 \ , \label{eqs_gen}
\eea
where the driving force is due to the logarithmic derivative of the potential $V$. 

A peculiar behavior emerges when these cosmologies are driven by exponential potentials, which can be parametrized as
\beq
    V(\varphi)\ = \ \frac{T}{2 \kappa^2}\ e^{2 \,\gamma\,\varphi}  \ . \label{Vgamma}
\eeq
In the orientifold models of~\cite{susy95,usp32}, which will be our main focus in this paper, these potentials would reflect an overall positive brane--orientifold tension $T$, which can be conveniently identified with the scale $M^2$ in eq.~\eqref{gauge_choice} and in the following, while $\gamma \geq 0$ characterizes the driving force on the scalar field due to $V(\phi)$~\footnote{The relevant values are $\gamma=\frac{5}{3}$ for the $SO(16) \times SO(16)$ model of~\cite{so1616}, where $T$ would reflect the one--loop vacuum energy, and $\gamma=1$ for the tachyon--free non--supersymmetric orientifolds of~\cite{susy95} and~\cite{usp32}.}. Note that in four dimensions, the exponential potential takes the form
\beq
\frac{M^2}{2 \kappa^2}\ e^{\kappa \, \gamma\, \sqrt{6} \, \phi} \ ,
\eeq
when expressed in terms of the canonically normalized field $\phi$.

The ``parametric time'' $\tau$ is related to the cosmic time according to
\beq
dt_c \ = \ \sqrt{\frac{2}{3}} \ e^{-\,\gamma\,\varphi} \ \frac{d\tau}{M} \ , \label{tctaugamma}
\eeq
and to the conformal time according to
\beq
d\eta \ = \ \sqrt{\frac{2}{3}} \ e^{-\, \frac{a}{3} \,-\, \gamma\,\varphi}\  \ \frac{d\tau}{M} \ . \label{etataugamma}
\eeq
With an exponential potential, the second of eqs.~\eqref{eqs_gen} reduces to
\beq
    \frac{d^2\varphi}{d \tau^2}\ +\ \frac{d\varphi}{d \tau}\sqrt{1\ + \ \left(\frac{d\varphi}{d \tau}\right)^2}\ + \ \gamma\left[1\ + \ \left(\frac{d\varphi}{d \tau}\right)^2\right] \ = \ 0\ ,
\eeq
and for $\gamma<1$ there are \emph{two} classes of expanding solutions, with quite different scalar dynamics.
\begin{itemize}
    \item[a.] In \emph{``descending'' cosmologies} the scalar field $\varphi$ emerges from the initial singularity from~\emph{large positive} values, and then decreases (one can thus say that $\varphi$ ``descends'' the potential) during the subsequent cosmological evolution. 

\item[b] In \emph{``climbing'' cosmologies} the scalar field emerges from~\emph{large negative} values of $\varphi$, and ``climbs up'' the potential until it reaches a turning point before it starts to descend. In String Theory, this solution has the virtue of involving a bounded string coupling, since $\varphi$ is naturally related to the dilaton~\cite{fss}.
\end{itemize}
 Both solutions eventually approach the limiting behavior described by
\beq
    \varphi(\tau)\ =\ \varphi_0 \ - \ \frac{\gamma \,\tau}{\sqrt{1-\gamma^2}}\ , \qquad a(\tau)\ =\ \frac{\tau}{\sqrt{1-\gamma^2}} \ ,
\eeq
which is the well--known Lucchin--Matarrese attractor~\cite{lm} in the present gauge. In conformal coordinates, this limiting behavior is described by
\beq
    ds^2 \ = \  \left[\frac{\sqrt{6(1-\gamma^2)}}{M(1-3\gamma^2)} \ \left(-\eta\right) \ e^{-\gamma \varphi_0}\right]^{\frac{2}{1-3\gamma^2}}\Big(-\ d\eta^2\ + \ d \vec{x} \cdot d\vec{x}\Big)\ , \eeq
while
\beq
\varphi(\eta) \ = \  \frac{\varphi_0}{1-3\gamma^2}\ + \ \frac{3\gamma}{1-3\gamma^2}\log \left[\frac{M(1-3\gamma^2)}{\sqrt{6(1-\gamma^2)}}(-\eta)\right] \ .
\eeq

Note that for $\gamma \ll 1$
\beq
   \varphi(\eta) \ \simeq \ \varphi_0 \ + \ 3 \gamma \,\log\left[\frac{M}{\sqrt{6}} \left(- \eta\right) \right]\ , \qquad e^{\frac{1}{3}\,a(\eta)} \ \simeq \ \sqrt{6} \ e^{-\gamma \varphi_0} \frac{1}{M \left(- \eta\right)} \ , \label{aphidS}
\eeq
where we have retained contributions up to ${\cal O}(\gamma)$,
so that when the system approaches the Lucchin--Matarrese attractor one is describing a de Sitter phase with
\beq
    {\cal H} \ =\ \frac{M}{\sqrt{6}} \ e^{\gamma \varphi_0} \label{HM}
     \ .
\eeq
For both classes of solutions the preceding constant value persists until $\tau={\cal O}\left(\frac{1}{\gamma^2}\right)$, and then ${\cal H}(\tau)$ starts to decrease. 

The preceding solutions give rise to power--law inflation for $\gamma< \frac{1}{\sqrt{3}} \simeq 0.58$, but the situation changes drastically as $\gamma \to 1$. Both the descending solution and the Lucchin--Matarrese attractor then disappear, while the leftover \emph{climbing} solution takes the simple form
\beq \varphi(\tau)\ = \ \widehat{\varphi}_0 \ + \ \frac{1}{2}\left(\log \tau \ - \ \frac{\tau^2}{2}\right) \ , \qquad a(\tau)\ = \ \frac{1}{2}\left(\log \tau \ + \ \frac{\tau^2}{2}\right)\ . \label{critical_beh}
\eeq

%%%%%%%%%%%%%%%%%%%%%%%%%%%%%%%%%%
\section[Climbing and the Primordial Power Spectrum]{\sc Climbing, \texorpdfstring{$\Delta$} \ \ and the Primordial Power Spectrum}\label{sec:climbing2}
%%%%%%%%%%%%%%%%%%%%%%%%%%%%%%%%%%

For $\gamma \geq 1$ only climbing solutions exist, but here we shall confine our attention to a hard exponential term with $\gamma=1$, which is directly related to the non--supersymmetric ten--dimensional orientifolds. This case will be central in our discussion, while the behavior for $\gamma < 1$ will be just indicative of what happens in slow-roll phases.

The climbing behavior is forced by the ``hard exponential'', and so it also occurs in the presence of more general potentials of the form
\beq
V \ = \ \frac{M^2}{2 \kappa^2}\ e^{2 \,\varphi} \ + \ v(\varphi) \ , \label{vmild}
\eeq
with $v(\varphi)$ a milder non--singular contribution, and can provide a reason for the onset of inflation. This is true, in particular, for the ``double--exponential potentials''
\beq
 V \ = \ \frac{M^2}{2 \kappa^2}\ e^{2 \,\varphi} \ + \ v_0 \,e^{2 \,\gamma\,\varphi} \ , \label{2exp}
\eeq
with $v_0>0$ and $\gamma< \frac{1}{\sqrt{3}}$.
Despite its simplicity, this setting is not exactly solvable, but the results recalled in the previous section capture its main features, since one can work with 
\beq
V \ = \ v_0 \,e^{2 \,\gamma\,\varphi} \label{2exp1}
\eeq
during most of the evolution, where the ``hard'' exponential has sub--dominant effects, and with
\beq
V \ = \ \frac{M^2}{2 \kappa^2}\ e^{2 \,\varphi}  \label{2exp2}
\eeq
near the turning point, where the ``hard exponential'' dominates. Integrable variants of the potential~\eqref{2exp} exist, and in particular the simple exact solution in the presence of
\beq
V \ = \ \frac{M^2}{2 \kappa^2}\left[ e^{\frac{2}{\gamma} \,\varphi} \ + \ e^{{2}{\gamma} \,\varphi}\right] \ . \label{integrable} 
\eeq
displays very neatly the transition from an early climbing phase to the onset of inflation~\cite{fss}.

\begin{figure}[ht]
\begin{center}
    \includegraphics[width=3.2in]{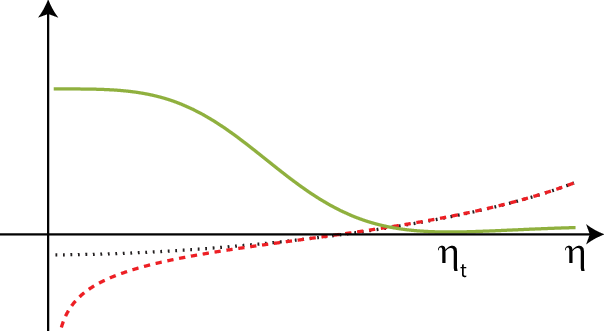}
\end{center}
\caption{Typically the scale factor (red, dashed) approaches the attractor (black, dotted) while the scalar field is still in the climbing phase with medium--large values of $\epsilon$ (green, solid), but an initial singularity is readily encountered at earlier times in the region where $\epsilon \simeq 3$. The dip in the solid curve for $\epsilon$ corresponds to the turning point.}
\label{fig:attractor}
\end{figure}

For the typical values $\gamma={\cal O}(0.1)$ that are selected by rough comparisons with CMB data, a de Sitter--like expansion takes place during the final descent. The results collected in~\cite{dkps} actually indicate that the same attractor behavior, with an essentially constant ${\cal H}$, also characterizes the final part of the earlier ascent. We shall thus assign to the scale factor of these systems a de Sitter--like behavior after the turning point, and in particular the de Sitter relation between cosmic and conformal time,
\beq
dt_c \ = \ - \ \frac{d\eta}{{\cal H} \,\eta} \ ,
\eeq
will be used repeatedly in the following. Details of the scalar dynamics play a key role near the turning point, where demanding an essentially constant value ${\cal H}$ for the Hubble parameter links ${\cal H}(\tau)$ to the microscopic parameters of the preceding section for the ``hard'' exponential. In that region $H$ evolves according to
\beq
{H}(\tau)\ = \ \frac{{M}_0}{2\,\sqrt{6}}\left(\frac{1}{\tau}+\tau\right)\tau^{1/2}\ e^{-\tau^2/4}\ e^{\varphi_0}  \ ,
\eeq
in terms of the parametric time $\tau$, 
where $M_0$ is the scale associated to the ``hard'' exponential, 
and identifying with ${\cal H}$ the value attained by $H(\tau)$ at $\tau=1$, which identifies the turning point, leads to
\beq
{\cal H} \ = \ \frac{M_0}{\sqrt{6}}\ e^{-\frac{1}{4}}e^{{\varphi}_0}  \ . \label{HM0}
\eeq
\begin{figure}[ht]
\begin{center}
    \includegraphics[width=3.3in]{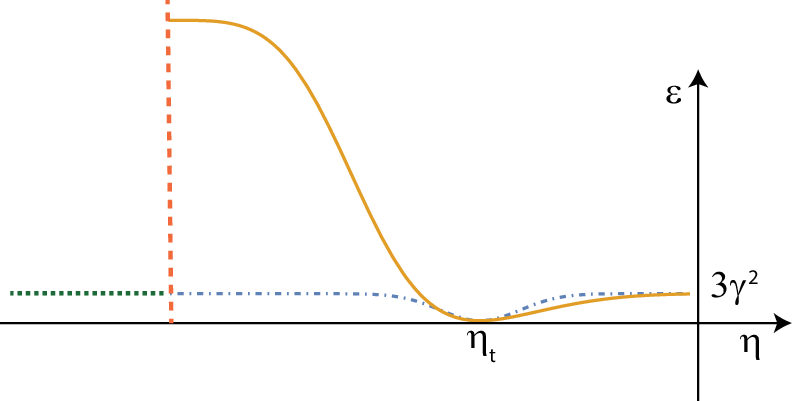}
\end{center}
\caption{The non--singular model for $\epsilon(\eta)$  (blue, dot-dashed), with a dip at $\eta_t$, where the climbing scalar inverts its motion. As $\eta$ decreases, the actual $\epsilon(\eta)$ (orange, solid) experiences a quick growth to its asymptotic value as the Universe collapses, and the dashed vertical line corresponds to the initial singularity.}
\label{fig:model_eps}
\end{figure}

A slow--roll phase is usually characterized by a set of slow--roll parameters, and primarily by
\beq
\epsilon \ = \ \frac{3\,\left(\frac{d\varphi}{d \tau}\right)^2}{1 \ +\  \left(\frac{d\varphi}{d \tau}\right)^2} \ , \qquad \delta \ = \ 3\,\frac{\frac{d^2\varphi}{d t_c^2}}{\frac{d\varphi}{d t_c} \, \frac{d a}{d t_c}} \ . \label{delta_epsilon}
\eeq
As can be clearly seen in Fig.~\ref{fig:attractor}, tracing the scalar dynamics backwards one encounters a zero of $\epsilon$ at $\eta_t$, which corresponds to the turning point, and a sharp increase at earlier times, which is accompanied by the sudden encounter with the initial singularity. To the best of our knowledge, there is no satisfactory way to deal with the divergent contribution that the singularity introduces in three-point amplitudes, but the higher--derivative corrections present in String Theory are expected to resolve it. We shall therefore assume, with good motivations but with an evident degree of arbitrariness, that the singularity would be resolved in a full string treatment, and that the growth of $\epsilon$ stops as it quickly recovers a slow--roll value, say $3 \gamma^2$ again, as sketched in fig.~\ref{fig:model_eps}. As we shall see, if this picture is somehow correct, the three--point amplitude places intriguing constraints on the earlier dynamics.

The climbing behavior typically introduces a low--frequency cut in the primordial power spectrum of curvature perturbations, which resonates with the apparent lack of power present in the first few CMB multipoles. This feature is generally accompanied by a peak that reflects the end of the climbing phase, whose position and size are, however, non--universal features of the scalar dynamics. Leaving aside these finer details, a simple formula,
\beq
P(k) \ = \ A \ \frac{\left(\frac{k}{k^\star}\right)^3}{\left[ \left(\frac{k}{k^\star}\right)^2 \ + \ \left(\frac{\Delta}{k^\star}\right)^2 \right]^{2 - \frac{n_s}{2}}} \label{power_delta} \ ,
\eeq
depending on a single additional parameter, the scale $\Delta$, aside from $A$ and the reference scale $k^\star$, can capture the gross features of the low--frequency cut while also approaching the Chibisov--Mukhanov tilt
\beq
P(k) \ \sim \ A \ \left(\frac{k}{k^\star}\right)^{n_s-1} \label{chib-muk}
\eeq
for values of $k$ above $\Delta$. 
The modified power spectrum in eq.~\eqref{power_delta} is actually determined by an exact solution of the Mukhanov--Sasaki equation 
\beq
    v''_k(\eta)\ +\ \left(k^2\ + \ \Delta^2 \ - \ \frac{\nu^2\ - \ \frac{1}{4}}{\eta^2}\right)v_k(\eta)\ = \ 0 \ ,
\eeq
with the potential
\beq
W \ = \ \frac{\nu^2\ - \ \frac{1}{4}}{\eta^2} \ - \ \Delta^2 \ , \label{W_Delta}
\eeq
which obtains lowering by $\Delta^2$ the standard attractor $W$.

Some evidence for the scale $\Delta$ was found in~\cite{gkmns} in the CMB data. The detection level improves up to the 3$\sigma$ level as the mask around the Galactic plane is enlarged, and the sharpest result,
\beq
\Delta \ = \ \left (0.351 \pm 0.114\right) \times 10^{- 3}\  \mathrm{Mpc}^{- 1} \ , \label{delta_value}
\eeq
is obtained within an open sky fraction $f_{sky}\simeq 39\%$. The preceding value is comparable to the Cosmic Horizon, and retracing the past history of the Universe one can translate it into an energy scale at the onset of inflation~\cite{gkmns}
\beq
\Delta_\mathrm{inf} \ \simeq \ 3 \times 10^{14} \ e^{N-60} \ \sqrt{\frac{H_\mathrm{inf}}{\mu_{\mathrm{Pl}}}} \ \mathrm{GeV} \ \simeq \ 2 \times 10^{12} \ \ e^{N-60} \ \mathrm{GeV} \ , \label{DeltaH}
\eeq
where $H_\mathrm{inf} \simeq 10^{14}\ \mathrm{GeV}$, $\mu_{\mathrm{Pl}} \simeq 2.4 \times 10^{18}\ \mathrm{GeV}$ is the reduced Planck energy and $N$ denotes the number of inflationary $e$-folds. 

%%%%%%%%%%%%%%%%%%%%%%%%%%%%%%%%%%
\section[\texorpdfstring{$\Delta$} \ \  and the Three--Point Amplitude]{\sc \texorpdfstring{$\Delta$} \ \  and the Three--Point Amplitude}\label{sec:Delta}
%%%%%%%%%%%%%%%%%%%%%%%%%%%%%%%%%%%%%

We can now link $\Delta$ to the conformal time of the turning point that concludes the climbing phase. To this end, let us start by recalling that the combination
\beq
z \ = \  e^\frac{a}{3}\ \sqrt{2 \,\epsilon} 
\eeq
satisfies
\beq
\frac{z''}{z} \ = \ W \ ,
\eeq
where $W$ is the Mukhanov--Sasaki potential, whose role was recalled in Section~\ref{sec:climbing2}.
During the descent, the scale factor has settled on the attractor value
\beq
e^{\frac{1}{3}\,a_{LM}}\ \sim \ (-\eta)^{-\frac{1}{1-3\gamma^2}} \ ,
\eeq
and, as we have stressed, this is typically the case already before the climbing phase ends. Under this assumption, one can deduce the behavior of
\beq
\sigma \ = \ \sqrt{\epsilon} 
\eeq
from the linear differential equation
\beq
        \sigma''\ + \ \left(1\,-\,2\nu\right)\frac{\sigma'}{\eta} \ + \ \Delta^2\,\sigma\ = \ 0 \ , \label{sigmadelta}
\eeq
where
\beq
\nu \ = \ \frac{3}{2} \frac{1 \ - \ \gamma^2}{1 \ - \ 3\,\gamma^2} \ .
\eeq

Demanding that $\epsilon$ approach $3\gamma^2$ as $\eta \rightarrow 0^-$, one can conclude that
    \newline
\beq
        \sigma(\eta)\  \simeq \ c_1 (-\Delta \eta)^{\nu} J_{\nu}(-\Delta \eta) \ - \  \frac{\pi}{2^{\nu}\Gamma(\nu)}\sqrt{\frac{2\nu-3}{2\nu-1}}(-\Delta \eta)^{\nu} Y_{\nu}(-\Delta \eta)\ . \label{sigma_delta}
\eeq
For negative values of $c_1$, the first zero $\eta_t$ of $\sigma$ approaches zero, while for large positive values it approaches $-\,\frac{4.5}{\Delta}$. We shall leave $\eta_t$ as a free parameter in the range $\left[-\,\frac{4}{\Delta},-\,\frac{1}{\Delta}\right]$, where the lower values correspond to stronger bounces of the scalar field against the ``critical'' exponential wall, while the higher values correspond to milder ones.
\begin{figure}[ht]
\centering
\begin{tabular}{ccc}
%\mbox{graphic} & \mbox{table} \\
\includegraphics[width=70mm]{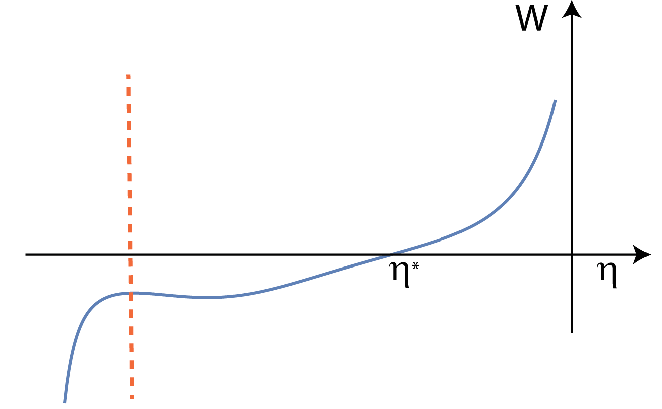} \quad  &
\includegraphics[width=70mm]{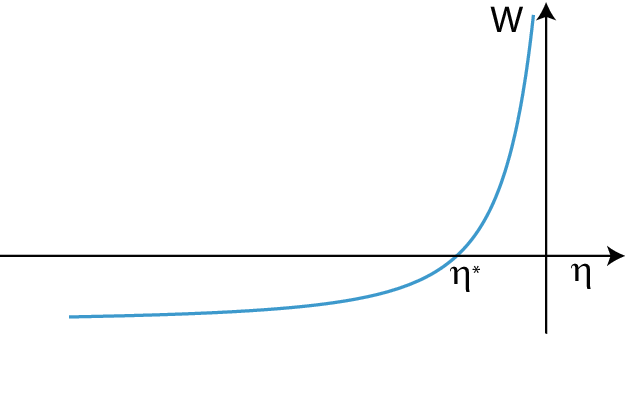}  \\
\end{tabular}
\caption{Left panel: a typical Mukhanov--Sasaki potential for a climbing scalar. Note that $W$ changes sign at $\eta^\star$ and experiences a short growth as $\eta$ decreases further, before tending to $-\infty$ as the initial singularity is approached. Right-panel: an attractor $W$ with $\nu=\frac{3}{2}$ lowered by $\Delta^2$, as in eq.~\eqref{W_Delta}.}
\label{fig:Ws}
\end{figure}

The evolution can be traced backward until the first zero of $\sigma$ or $\epsilon$, which can be identified with the turning point at $\eta_t$ and links it to the scale $\Delta$. This result will play an important role in our construction of the three--point amplitude, but now the issue is what happened earlier.
Fig.~\ref{fig:Ws} compares the typical behavior of $W$ for a climbing--scalar cosmology with an attractor $W$ lowered by $\Delta^2$ as in eq.~\eqref{W_Delta}. 

The simplified shape in the right panel of fig.~\ref{fig:Ws} suffices to model the low-$k$ behavior of the primordial power spectrum and, as we have seen, can also recover the presence of a turning point. 
However, the three--point amplitude is built from contributions of the type
\beq
\int_{-\infty}^0 d\eta \,f(\eta) \Big[G_1(k_1,\eta)\, G_2(k_2,\eta)\, G_3(k_3,\eta) \ - \ \mathrm{c.c.}  \Big] \ , \label{3pointcc}
\eeq
where the $G$'s denote Green functions or their first time derivatives, and depends crucially on the earlier history of the system, so more information is needed to compute it. One could try to continue with the actual $W$ determined by the two--derivative Lagrangian. However, the scalar moves faster and faster for earlier times, so that $\epsilon$ quickly approaches three, its upper bound, and one encounters an initial singularity at a nearby finite value $\eta_s$ of the conformal time, where the Universe collapses. This fate should be avoided in a complete theory of gravity, and, most importantly for us, the three--point amplitude suffers there from what one could call an \emph{ultraviolet divergence}, since some integrands are dominated by
\beq
\frac{\log(\eta\,-\,\eta_s)}{\eta \,-\, \eta_s} ,
\eeq
as $\eta \to \eta_s^+$. This is a problem if one tries to enforce the Bunch--Davis condition there, where the evolution determined by eq~\eqref{2der_act} starts.
\begin{figure}[ht]
\centering
\begin{tabular}{ccc}
%\mbox{graphic} & \mbox{table} \\
\includegraphics[width=65mm]{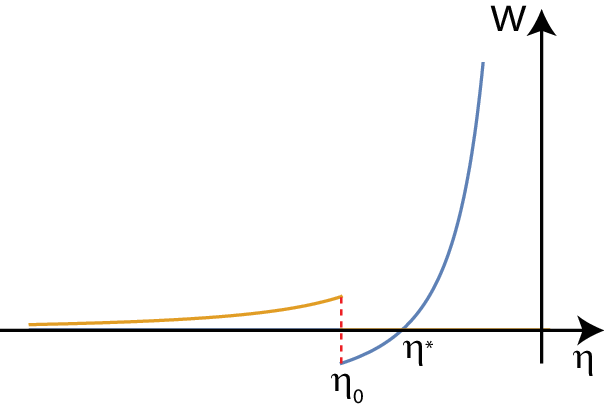} \quad  &
\includegraphics[width=65mm]{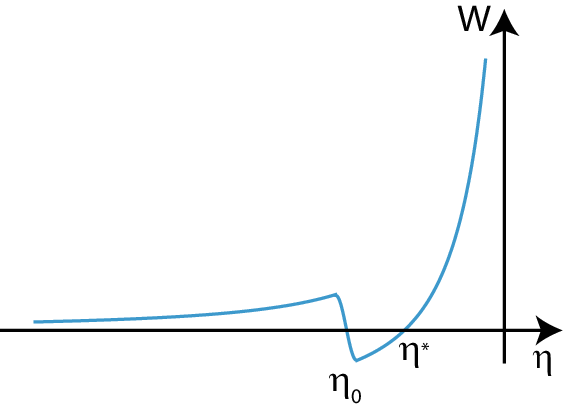}  \\
\end{tabular}
 \caption{\small Left panel: an extended model for $W$ that follows the curve of eq.~\eqref{W_Delta} up to $\eta_0$, where it experiences a sudden jump to the corresponding attractor curve without $\Delta$, has the virtue of leading to simple mode functions. Right panel: the actual transition could occur smoothly, within a small window of conformal time.}
\label{fig:Wsm}
\end{figure}

Trying to avoid this problem leads to an interesting scenario, as we can now explain.
String Theory is supposed to regulate the divergence, albeit in ways that we do not control at present, and therefore some guesswork is needed. One could conceive, for example, that the higher--derivative corrections drive $\epsilon$ to recover its slow--roll value $3 \gamma^2$, as in fig.~\ref{fig:model_eps}, while the Universe continues a de Sitter contraction. If this were the case, the two--derivative equations would suggest that $W$ returns to the attractor form without $\Delta$, as in fig.~\ref{fig:Wsm}, so that $\epsilon$ could stop moving. Note, in fact, that only if $\Delta=0$ is eq.~\eqref{sigma_delta} solved by a constant $\sigma$, so that $\epsilon$ could stay anchored to the attractor value as $\eta$ decreases beyond the turning point, consistent with the resolution of the singularity illustrated in fig.~\ref{fig:model_eps}. 
This picture can also lead to a simple model, where $\Delta$ disappears suddenly as in the left panel of fig.~\ref{fig:Wsm}, and the mode functions can be computed exactly. However, the reader will recognize that, as in one--dimensional Quantum Mechanics, a Bunch--Davies mode function $H_{3/2}^{(1)}$ on the left would call for contributions involving both $H_{3/2}^{(1)}$ and $H_{3/2}^{(2)}$ on the right. 
The presence of both types of mode functions would introduce in the three-point amplitude what could be termed an \emph{infrared divergence}, since the problem now arises from the upper end of the integration region. This divergence could not be compensated for by the field redefinitions, which yield only finite results, and would also affect higher--point functions, even if further redefinitions, as in~\cite{maldacena}, were used to remove terms without time derivatives from the three--point amplitude.
\begin{figure}[ht]
\centering
\begin{tabular}{ccc}
%\mbox{graphic} & \mbox{table} \\
\includegraphics[width=70mm]{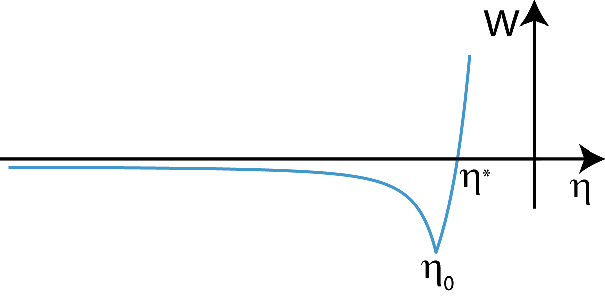} \quad  &
\includegraphics[width=70mm]{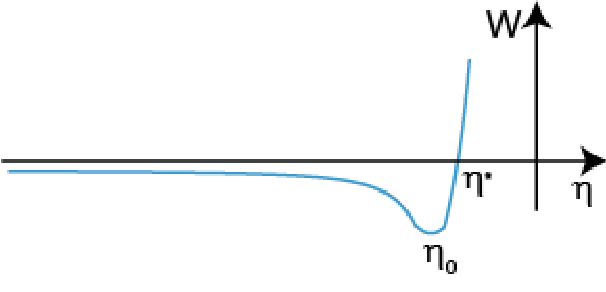}  \\
\end{tabular}
 \caption{\small Left panel: a sharp transition between an earlier epoch of compression and the expansion that begins at $\eta_0$ and evolves into the inflationary scenario. Right panel: a smooth model of the transition.}
\label{fig:Wsm2}
\end{figure}

We can now explore an alternative option that can avoid these difficulties. It is suggested by the growth experienced by $W$ in fig.~\ref{fig:Ws} for decreasing values of $\eta$, before the initial singularity takes over. What if this behavior persisted in a full string treatment, as in fig.~\ref{fig:Wsm2}? If $\epsilon$ quickly recovered a slow--roll value, say again $3\,\gamma^2$, the two--derivative equations would compel the Universe to expand again for decreasing values of $\eta$. The initial singularity would therefore be replaced by a bounce at a conformal time $\eta_0$ below $\eta_t$, both typically ${\cal O}\left(-\, \frac{1}{\Delta}\right)$, and the earlier history of the Universe would play no role in the subsequent evolution. 

If a bounce were induced at $\eta_0$ by string corrections, the Bunch--Davies condition should naturally be imposed there, at the instant of maximal compression. One might object that the same inflationary evolution would emerge if one started in a Bunch--Davies vacuum near the end of the climbing phase, without appealing to string--induced modifications of the action in eq.~\eqref{2der_act}. However, the bounce scenario grants this peculiar choice of initial condition a special role and is somehow suggested by the typical $W$ in the left panel of fig.~\ref{fig:Ws} before the singularity is reached. Moreover, the assumption that the singularity is somehow overcome by string corrections conforms to standard expectations, and is not foreign to other treatments of non-Gaussianities that we are aware of, starting from the original work of~\cite{maldacena}, where the de Sitter phase continues indefinitely in the past. 

A bounce scenario was actually considered long ago in String Theory by Gasperini and Veneziano~\cite{gasp_ven}, under the spell of $T$-dualities. It resolves most of the problems connected with our setting, so we shall now explore its consequences, although it creates another problem. In fact, the reader will not fail to recognize that the modifications impinge on the unique role of the climbing behavior, which will continue to underlie our considerations but, strictly speaking, is only unavoidable in the two--derivative setting.

After the bounce, the Universe undergoes a slow--roll evolution that is similar to the scenario originally considered in~\cite{maldacena} and reviewed in~\cite{reviews}, but for three important distinctive features:
\begin{itemize}
    \item the attractor curve for $W$ is lowered by $\Delta^2$, consistently with the modified power spectrum of eq.~\eqref{power_delta}, so that the mode functions depend on
        \beq
\omega \ = \ \sqrt{k^2 \ + \ \Delta^2} \ ; \label{omegakappa}
    \eeq
    \item the expected contribution to the three--point amplitude acquires an oscillatory behavior depending on $\omega$
    and on the conformal time $\eta_0$ where the expansion begins, which as we stated is typically ${\cal O}\left(-\, \frac{1}{\Delta}\right)$;
    \item \emph{curvature} perturbations are actually singular at the turning point $\eta_t$ but, as we shall see, in a way that is regulated by the Schwinger--Keldysh contour. The resulting contributions have an oscillatory behavior determined by $\omega$ and $\eta_t$.
\end{itemize}
This final setting seems relatively simple, and the analysis we are about to present seems well motivated and safe, but we are aware that subtleties of various types were considered in the literature, in a number of different contexts, in particular in~\cite{sub}. 
In our case, the  solutions of the Mukhanov--Sasaki equation rest essentially on Bessel functions of order $\frac{3}{2}$, and for $\eta>\eta_0$ the initial Bunch--Davis condition selects the mode functions
\beq
    v_\omega(\eta)\ =  \ - \ \frac{\sqrt{\pi}}{2}\sqrt{-\eta}\ H_{\frac{3}{2}}^{(1)}(-\omega \eta) \ = \  \frac{1}{\sqrt{2 \,\omega}} \left(1 \ - \ \frac{i}{\omega\,\eta} \right)\ e^{-i\,\omega\,\eta} \ , \label{vlargenegeta}
\eeq
with $\omega$ as in eq.~\eqref{omegakappa}.
Consequently, for $\eta > \eta_0$ the Green functions that determine the three--point amplitude are
\beq
G_>(\eta,\omega) \ = \ \frac{{\cal H}^2}{4 \,M_P^2 \, \sqrt{3 \, \gamma^2 \, \epsilon (\eta)} \ \omega^{3}} \ e^{i \, \omega\,\eta} \left(1\ - \ i\, \omega\,\eta \right)  \ ,
\label{green}
\eeq
with $\epsilon(\eta) = 3 \,\gamma^2$ away from the turning point.
\begin{figure}[ht]
\begin{center}
    \includegraphics[width=3in]{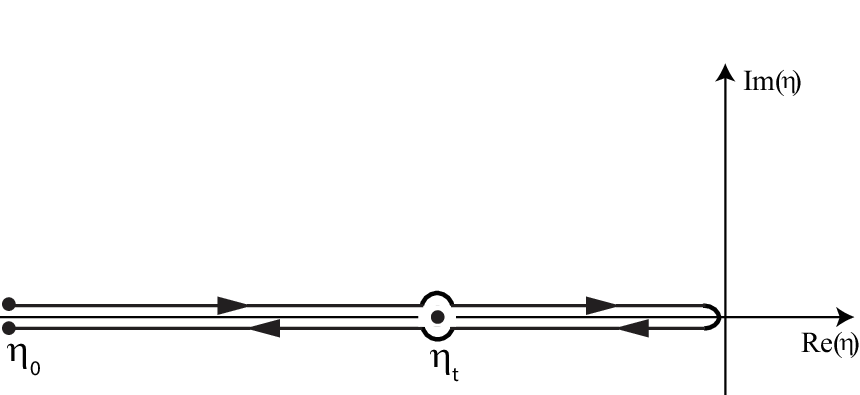}
\end{center}
\caption{The Schwinger--Keldysh contour for the proposed dynamics starts slightly above the real axis of the $\eta$ plane at $\eta_0$, proceeds along the semicircle above $\eta_t$, turns back at the origin, proceeds along the semicircle below $\eta_t$ and finally ends at $\eta_0$ slightly below the real axis.}
\label{fig:contour}
\end{figure}

The three-point amplitude thus rests on three types of ingredients, which depend on the two conformal times $\eta_0$ and $\eta_t$, both ${\cal O}\left(-\,\frac{1}{\Delta}\right)$ and with $\eta_t > \eta_0$ in our picture:
\begin{enumerate}
    \item a ``principal--part'' contribution from the region $\eta_0 \leq \eta< 0$, $\eta \neq \eta_t$, on a Schwinger--Keldysh contour whose two sides are slightly displaced, as fig.~\ref{fig:contour}, around the real axis, where the Green function is as in eq.~\eqref{green} with  $\epsilon(\eta)=3\,\gamma^2$;
    \item a contribution emerging from the neighborhood of $\eta_t$, the turning point where $\epsilon(\eta)$ vanishes, which is regulated by the Schwinger--Keldysh contour;
    \item a contribution arising from the field redefinition performed in~\cite{maldacena} and reviewed in~\cite{reviews}, which is identical to the original result, up to the replacement of $k_i$ with $\omega_i$.
\end{enumerate}

%%%%%%%%%%%%%%%%%%%%%%%%%%%%%%%%%%
\section[The Three--Point Amplitude]{\sc  The Three--Point Amplitude}\label{sec:3point}
%%%%%%%%%%%%%%%%%%%%%%%%%%%%%%%%%%%%%
We can now compute the three--point amplitude, proceeding as outlined in the preceding section. 
The cubic action derived in~\cite{maldacena} and reviewed in detail in~\cite{reviews} reads
\bea
    \mathcal{S}^{(3)} &=&  M_{P}^2\int d^4 x \Big\{\epsilon^2 e^{a}\dot{\zeta}^2\zeta\,+\,\epsilon^2 e^{a/3}\zeta \partial_k \zeta \partial^k \zeta \,-\,2 \epsilon^2 e^{a} \dot{\zeta} \partial_k \zeta \partial^k (\partial^{-2} \dot{\zeta})\nonumber \\ 
    &+&  \frac{3\,\epsilon}{2}\, e^{a}\,\frac{d}{dt_c}\left(\frac{\dot \epsilon}{\epsilon \dot a}\right)\dot{\zeta}\zeta^2\,-\,\frac{\epsilon^2}{2}\ e^{a} \left[\dot{\zeta}^2\zeta - \zeta \partial_k \partial_l (\partial^{-2}\dot{\zeta})\partial^k \partial^l (\partial^{-2}\dot{\zeta})\right]+f(\zeta)\frac{\delta \mathcal{L}}{\delta \zeta}\Big\}\,,
\eea
where ``dots'' denote derivatives with respect to cosmic time,
and where the last term contains the field redefinition originally performed in~\cite{maldacena}, which emphasizes the slow--roll limit of the different residual contributions, with
\bea
    f(\zeta)&=&(\delta + \epsilon)\zeta^2 \ + \ \frac{6}{\dot{a}}\,\dot{\zeta}\zeta \ - \ \frac{9}{2\,\dot{a}^2}\ e^{-\frac{2}{3}a}\left[\partial_k\zeta \partial^k \zeta - \partial^{-2}\partial_k \partial_l (\partial^k \zeta \partial^l \zeta)\right]\nonumber \\ 
    &+&\frac{3\,\epsilon}{\dot{a}}\ \left[\partial_k \zeta \partial^k (\partial^{-2}\dot{\zeta})\,-\,\partial^{-2}\partial_k \partial_l (\partial^k \zeta \partial^l (\partial^{-2} \dot{\zeta}))\right] \ .
\eea
Following the convenient classification in~\cite{reviews}, we can now write the three--point amplitude as
\bea
    {\cal A}_3 &=& \int \frac{d^3 \vec k_1}{(2 \pi)^3} \frac{d^3 \vec k_2}{(2 \pi)^3} \frac{d^3 \vec k_3}{(2 \pi)^3}e^{i \vec k_1 \cdot \vec x} e^{i \vec k_2 \cdot \vec y} e^{i \vec k_3 \cdot \vec z}(2 \pi)^3 \delta^{(3)}(\vec k_1 + \vec k_2 + \vec k_3)\nonumber \\
    && \Big\{\langle O_1 \rangle + \langle O_2 \rangle + \langle O_3 \rangle + \langle O_4 \rangle + \langle O_5 \rangle + \langle O_f \rangle +  2\, \mathrm{perms} \Big\}\  ,
\eea
where the last term,
\bea
\langle {\cal O}_f \rangle &=& \frac{\mathcal{H}^4}{16\,M_P^4} \frac{\delta + \epsilon}{\epsilon^2} \, \frac{1}{ \omega_2^3 \, \omega_3^3} \ ,
\eea
with $\delta$ and $\epsilon$ as in eq.~\eqref{delta_epsilon},
originates from the field redefinition in~\cite{maldacena}, and, as we saw in Section~\ref{sec:climbing2},
\beq
\omega_{2,3} \ = \ \sqrt{k_{2,3}^2 \ + \ \Delta^2} \ ,
\eeq
since in our case the attractor curve is lowered by $\Delta^2$.

As we have anticipated, the three--point amplitude is determined by two contributions to the Schwinger--Keldysh contour displayed in fig.~\ref{fig:contour} with very different origins. Those of the first type, which we shall refer to as ``principal--part'' contributions, reflect the quasi--de Sitter expansion of the Universe with a constant $\epsilon=3 \,\gamma^2$ that begins at $\eta_0$ and leave out the turning point at $\eta_t$, while those of the second type emerge precisely from the turning point. Both types of contributions originate from the five integrals 
\bea
    \langle O_1 \rangle &=&  2i M_{P}^2 \int_{ \eta_0}^{0} \frac{d \eta\ \epsilon(\eta)^2}{(\mathcal{H} \eta)^2}\left[\, G_>'(1)\, G_>'(2) \,G_>(3)\  - \mathrm{c.c.}\right] \ , \nonumber \\
    \langle O_2 \rangle &=&  -\ 2i M_{P}^2 \int_{ \eta_0}^{0} \frac{d \eta\ \epsilon(\eta)^2}{(\mathcal{H} \eta)^2}\left(\vec k_1 \cdot \vec k_2\right)\left[ G_>(1)\, G_>(2) \,G_>(3)\  - \mathrm{c.c.} \right] \ ,  \nonumber \\
    \langle O_3 \rangle &=&  -\ 2i M_{P}^2 \int_{ \eta_0}^{0} \frac{d \eta\ \epsilon(\eta)^2}{(\mathcal{H} \eta)^2}\left[\frac{\vec k_1 \cdot \vec k_3}{k_1^2} +  \frac{\vec k_2 \cdot \vec k_3}{k_2^2}\right]\left[ \, G_>'(1) \, G_>'(2) \,G_>(3)\  - \mathrm{c.c.}\right] \ , \nonumber \\
    \langle O_4 \rangle &=&  i M_{P}^2 \int_{ \eta_0}^{0} \frac{d \eta}{(\mathcal{H} \eta)^2} \, \frac{\epsilon (\eta) }{\mathcal{H}}\,\frac{d^2}{dt_c^2}\log \epsilon(\eta)\, \left[ \, G_>'(1)\, G_>(2) \,G_>(3)\  - \mathrm{c.c.}\right]\ ,  \nonumber \\
    \langle O_5 \rangle &=&  -\ i M_{P}^2 \int_{ \eta_0}^{0} \frac{d \eta\ \epsilon(\eta)^3}{(\mathcal{H} \eta)^2} \, \left[1 \, - \, \frac{\left(\vec{k}_2\cdot \vec{k}_3\right)^2}{k_2^2\,k_3^2}\right]\left[ \, G_>'(1) \, G_>'(2) \,G_>(3)\  - \mathrm{c.c.}\right] \ . \label{O15}
\eea
In these expressions, the scale factor $e^\frac{a}{3}$ has been replaced by its de Sitter form $\frac{1}{- {\cal H}\,\eta}$, and the measure has been similarly adapted to the conformal time $\eta$. Note that $O_2$ involves products of three Green functions, while in $O_1$, $O_3$ and $O_5$ the two ``primed'' factors are differentiated with respect to the conformal time $\eta$. In $O_4$, which has no ``principal--part'' contribution since it contains derivatives of $\epsilon$, only one of the Green functions is differentiated.

%%%%%%%%%%%%%%%%%%%%%%%%%%%%%%%%%%
\subsection[Principal--Part Contributions]{\sc  Principal--Part Contributions}\label{sec:3point1}
%%%%%%%%%%%%%%%%%%%%%%%%%%%%%%%%%%%%%

In this section, we collect the ``principal--part'' contributions, showing for brevity only those for the equilateral configuration, which yields the largest result. These follow from standard integrations, but the reader should note that the $H_\frac{3}{2}^{(1)}$ behavior of the mode functions, without other terms involving $H_\frac{3}{2}^{(2)}$, is crucial to avoid a divergent contribution to $\langle O_2 \rangle$ from the upper end of the integration region. The results for the different amplitudes are
\bea
\langle O_1 \rangle &=&  \frac{\mathcal{H}^4}{144 \, \gamma^2 \, {M_p}^4 \omega^6}\ \Big[4 \left(1 \ - \ \cos (3 \eta_0 \, \omega )\right) \ - \ 3 \,\eta_0 \, \omega \sin (3 \eta_0 \,\omega )  \Big] \ , \nonumber\\
\langle O_2 \rangle &=& \frac{\mathcal{H}^4\ k^2}{288 \,\gamma^2 \, \eta_0 \, {M_p}^4 \omega^9}   \Big[3 \left(\eta_0^2 \,\omega^2\,-\,3\right) \sin (3 \eta_0 \, \omega )\ + \ \eta_0 \, \omega  \left(10 \cos (3 \eta_0 \,\omega )\,+\,17\right)\Big] \ ,\nonumber\\
\langle O_3 \rangle &=&  \frac{\mathcal{H}^4}{144 \, \gamma^2 \, {M_p}^4 \omega^6}\ \Big[4 \left(1 \, - \, \cos (3 \eta_0 \, \omega )\right) \ - \ 3 \,\eta_0 \, \omega \sin (3 \eta_0 \,\omega )  \Big] \ , \nonumber\\
\langle O_4 \rangle &=&  0 \ , \nonumber\\
\langle O_5 \rangle &=&  - \ \frac{\mathcal{H}^4}{128 \, {M_p}^4 \omega^6} \   \Big[4 \left(1 \, - \, \cos (3 \eta_0 \, \omega )\right) \ - \ 3 \,\eta_0 \, \omega \sin (3 \eta_0 \,\omega )  \Big] \ , \label{easy_contributions}
\eea
where, as we saw in Section~\ref{sec:climbing2},
\beq
\omega \ = \ \sqrt{k^2 \ + \ \Delta^2} \ .
\eeq

The contribution from $\langle O_4 \rangle$ vanishes identically, since it involves derivatives of $\epsilon$, which is constant away from the turning point in our approximation, while the others are proportional in the equilateral configuration. $\langle O_5 \rangle$ is subdominant in $\gamma$ or $\epsilon$, and for this reason it was ignored in~\cite{maldacena}. All these contributions oscillate around the result in~\cite{maldacena}, which can be recovered if one ignores the trigonometric functions.

%%%%%%%%%%%%%%%%%%%%%%%%%%%%%%%%%%
\subsection[Turning--Point Contributions]{\sc  Turning--Point Contributions}\label{sec:3point2}
%%%%%%%%%%%%%%%%%%%%%%%%%%%%%%%%%%%%%

We can now describe the contributions that originate from the neighborhood of the turning point, where the dynamics is generally dominated by the hard exponential. Again, for brevity, we only show the equilateral results. 
In the neighborhood of the turning point, the integration variable $\eta$ can be traded for the parametric time $\tau$, and one can rely on the explicit forms of $\varphi(\tau)$ and $a(\tau)$ in eq.~\eqref{critical_beh}, so that
\beq
\epsilon(\tau) \ = \ 3 \left(\frac{1 \ - \ \tau^2}{1 \ + \ \tau^2}\right)^2 \ .
\eeq
The microscopic data of the ``critical'' climbing scalar can be then linked to $\cal {H}$ as explained in Sections~\ref{sec:climbing} and~\ref{sec:climbing2}, and in particular making use of eq.~\eqref{HM0}.
In the two branches of the Schwinger--Keldysh contour, these local contributions are computed relying on
    \beq
\frac{1}{z \ \pm \ i\,\xi} \ = \ PP\left[\frac{1}{z} \right] \ \mp \ i\, \pi \, \delta(z) \ ,
    \eeq
and on the residues that emerge from the two portions of the integrands, which have individually (anti) analytic properties there.

There is a subtlety here: the Mukhanov--Sasaki variable and the deviation of the scalar field from its cosmological trajectory are connected by coordinate transformations that are singular at the turning point, so that they are somehow inequivalent in the case under scrutiny. With the former choice and thus focusing on \emph{curvature} perturbations, which are closer to actual experiments, the turning point leaves tangible and calculable imprints in the three--point amplitude, whose interesting features we can now describe.

Only three of the five amplitudes in eqs.~\eqref{O15} play a role, since the others are not singular at the turning point, and the resulting equilateral contributions read
\bea
\langle O_1 \rangle &=& -\ \frac{3\,\pi\, \mathcal{H}^5 \,{e}^\frac{1}{12}}{16\, \gamma^3 \,\eta_t^2\, {M_p}^4 \omega^9} \  \Big[\eta_t \omega  \left(\eta_t^2 \omega ^2-3\right) \sin (3 \eta_t \omega )\ + \ \left(3 \eta_t^2 \omega ^2-1\right) \cos (3 \eta_t \omega )\Big]  \ , \nonumber\\
\langle O_2 \rangle &=&  0 \ ,\nonumber\\
\langle O_3 \rangle &=&  -\ \frac{3\,\pi\, \mathcal{H}^5 \,{e}^\frac{1}{12}}{16\, \gamma^3 \,\eta_t^2\, {M_p}^4 \omega^9} \  \Big[\eta_t \omega  \left(\eta_t^2 \omega ^2-3\right) \sin (3 \eta_t \omega )\ + \ \left(3 \eta_t^2 \omega ^2-1\right) \cos (3 \eta_t \omega )\Big] \ , \nonumber\\
\langle O_4 \rangle &=&  \frac{\pi  \mathcal{H}}{288 \,{e}^\frac{1}{4} \gamma ^3 \eta_t^6 {M_p}^4 \omega ^9} \Big\{\eta_t \omega  \sin (3 \eta_t \omega ) \Big[\eta_t \Big(9 \eta_t^3 \omega ^4-65 \eta_t \omega^2-90 {e}^\frac{1}{4} \eta_t^2 \mathcal{H}^3 \left(\eta_t^2 \omega ^2-3\right)\nonumber \\ &+& 2 {e}^\frac{1}{6} \eta_t \mathcal{H}^2 \left(\eta_t^2 \omega ^2+15\right) \ + \ 9 {e}^\frac{1}{12} \mathcal{H} \left(\eta_t^4 \omega ^4-5 \eta_t^2 \omega ^2+4\right)\Big)\ + \ 60\Big]\nonumber \\
&+&\cos (3 \eta_t \omega ) \Big[\eta_t \Big(33 \eta_t^3 \omega^4-81 \eta_t \omega^2+90 {e}^\frac{1}{4} \eta_t^2 \mathcal{H}^3 \left(1-3 \eta_t^2 \omega ^2\right)\nonumber \\ &-& 2 {e}^\frac{1}{6} \eta_t \mathcal{H}^2 \left(3 \eta_t^4 \omega^4+12 \eta_t^2 \omega^2-5\right)\ + \ 3 {e}^\frac{1}{12} \mathcal{H} \left(9 \eta_t^4 \omega ^4-17 \eta_t^2 \omega ^2+4\right)\Big) \ +\ 20\Big]\Big\} \ , \nonumber\\
\langle O_5 \rangle &=& 0 \ . \label{turn_contributions}
\eea
Note that these contributions include terms accompanied by different powers of $\mathcal{H}$. In view of eq.~\eqref{DeltaH}, this has an interesting effect: it introduces a strong dependence of the three--point amplitude, and of the corresponding observables, on the number of inflationary $e$-folds.

%%%%%%%%%%%%%%%%%%%%%%%%%%%%%%%%%%
\section[Equilateral \texorpdfstring{$f_{NL}(k)$} \ \ and Comparisons]{\sc  Equilateral \texorpdfstring{$f_{NL}$} \ \ and Comparisons}\label{sec:fNL}
%%%%%%%%%%%%%%%%%%%%%%%%%%%%%%%%%%%%%
\begin{figure}[ht]
\centering
\begin{tabular}{ccc}
%\mbox{graphic} & \mbox{table} \\
\includegraphics[width=70mm]{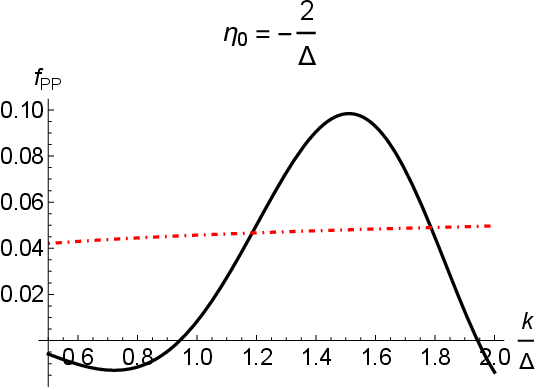} \quad  &
\includegraphics[width=70mm]{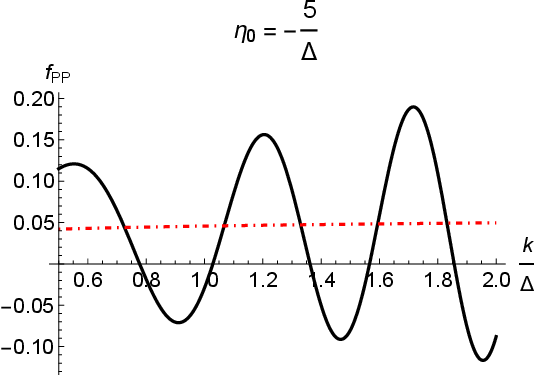}  \\
\end{tabular}
 \caption{\small Left panel: $f_{PP}$ (black, solid) compared with $f_{NL}^{(M)}$ (red, dot-dashed), for $\eta_0= -\,\frac{2}{\Delta}$. Right panel: $f_{PP}$ (black, solid) compared with $f_{NL}^{(M)}$ (red, dot-dashed), for $\eta_0= -\,\frac{5}{\Delta}$. The results displayed correspond to the choice $\epsilon=0.03$, and the oscillations are determined by $\eta_0$.}
\label{fig:fpp}
\end{figure}

Our starting point is the definition of the convenient dimensionless quantity
\beq
    f_{NL}(k)\ = \ 40 \, \frac{\gamma^4 \, M_{p}^4\,\Delta^{6}}{ \, \mathcal{H}^4}  \left[\left(\frac{k}{\Delta}\right)^2 + 1\right]^{2 \, \nu} \, \langle \zeta(k) \zeta(k) \zeta(k) \rangle \ ,
    \label{eq:fnl}
\eeq
where we have identified the reference scale $k^\star$ with $\Delta$ and
\beq
\langle \zeta(k) \zeta(k) \zeta(k) \rangle \ = \ \langle O_1 \rangle + \langle O_2 \rangle + \langle O_3 \rangle + \langle O_4 \rangle + \langle O_5 \rangle + \langle O_f \rangle +  2\, \mathrm{perms} \ .
\eeq
We also write
\beq
f_{NL}(k) \ = \ f_{PP}(k) \ + \ f_{t}(k) \ , \label{ppt}
\eeq
and in the following we consider separately the two contributions
of Sections~\ref{sec:3point1} and \ref{sec:3point2}, comparing the former with the original result in~\cite{maldacena}, which in the same conventions reads
\beq
    f_{NL}^{(M)}(k)\ = \ \frac{5}{12}\, \left(\frac{k}{\Delta}\right)^{4 \, \nu - 6} \left[-\,3\, \gamma^4\,+\,17\, \gamma^2\,+\,2\, \delta \right] \ .
    \label{eq:fnlM}
\eeq
\begin{figure}[ht]
\centering
\begin{tabular}{ccc}
%\mbox{graphic} & \mbox{table} \\
\includegraphics[width=70mm]{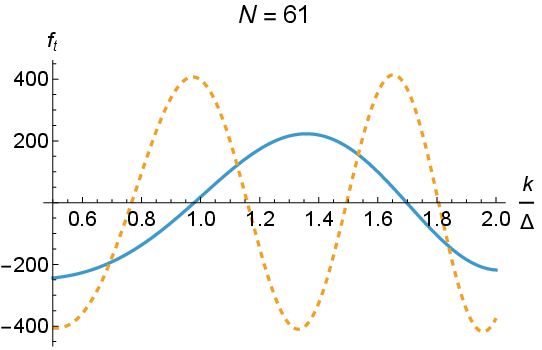} \quad  &
\includegraphics[width=70mm]{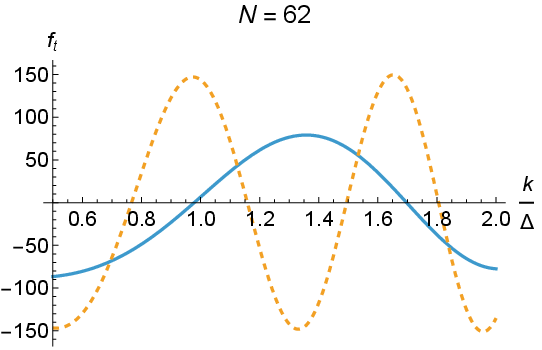}  \\
\end{tabular}
 \caption{\small Left panel: $f_{t}$ for $\eta_t= -\,\frac{2}{\Delta}$ (blue, solid) and for $\eta_t= -\,\frac{4}{\Delta}$ (orange, dashed) for $N=61$. Right panel: $f_{t}$ for $\eta_t= -\,\frac{2}{\Delta}$ (blue, solid) and for $\eta_t= -\,\frac{4}{\Delta}$ (orange, dashed) for $N=62$.  The results displayed correspond to the choice $\epsilon=0.03$, and the oscillations are determined by $\eta_t$.}
\label{fig:ft1}
\end{figure}
\begin{figure}[ht]
\centering
\begin{tabular}{ccc}
%\mbox{graphic} & \mbox{table} \\
\includegraphics[width=70mm]{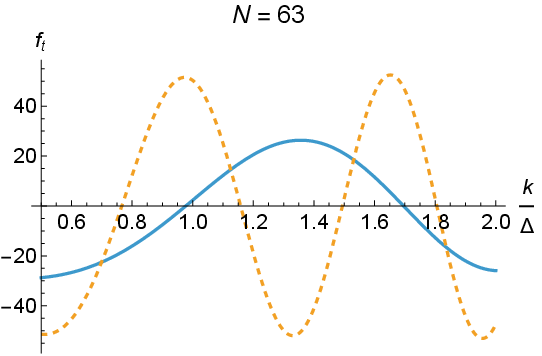} \quad  &
\includegraphics[width=70mm]{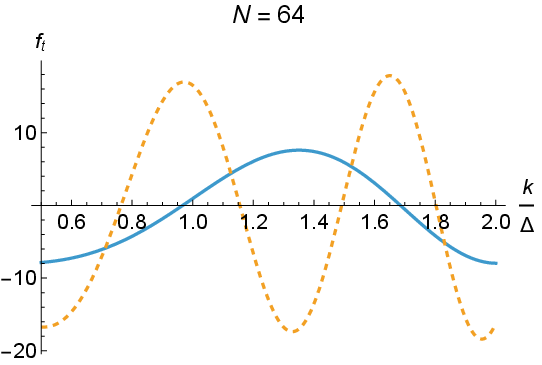}  \\
\end{tabular}
 \caption{\small Left panel: $f_{t}$ for $\eta_t= -\,\frac{2}{\Delta}$ (blue, solid) and for $\eta_t= -\,\frac{4}{\Delta}$ (orange, dashed) for $N=63$. Right panel: $f_{t}$ for $\eta_t= -\,\frac{2}{\Delta}$ (blue, solid) and for $\eta_t= -\,\frac{4}{\Delta}$ (orange, dashed) for $N=64$.  The results displayed correspond to the choice $\epsilon=0.03$, and the oscillations are determined by $\eta_t$.  }
\label{fig:ft2}
\end{figure}
\begin{figure}[ht]
\centering
\begin{tabular}{ccc}
%\mbox{graphic} & \mbox{table} \\
\includegraphics[width=70mm]{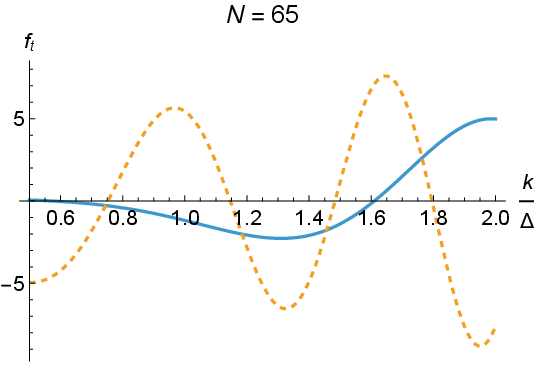} \quad  &
\includegraphics[width=70mm]{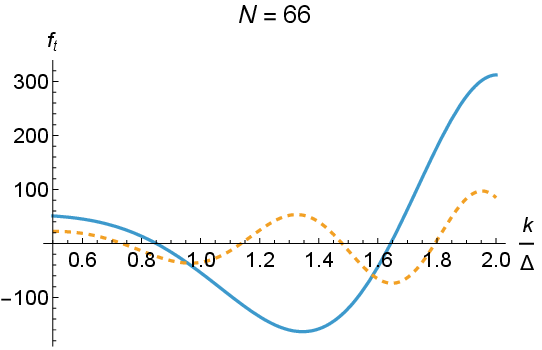}  \\
\end{tabular}
 \caption{\small Left panel: $f_{t}$ for $\eta_t= -\,\frac{2}{\Delta}$  (blue, solid) and for $\eta_t= -\,\frac{4}{\Delta}$ (orange, dashed) for $N=65$. Right panel: $f_{t}$ for $\eta_t= -\,\frac{2}{\Delta}$  (blue, solid) and for $\eta_t= -\,\frac{4}{\Delta}$ (orange, dashed) for $N=66$.  The results displayed correspond to the choice $\epsilon=0.03$, and the oscillations are determined by $\eta_t$.  }
\label{fig:ft3}
\end{figure}
Here we include for completeness the subdominant term contributed by $\langle O_5 \rangle$, which was ignored in the original work. This result further simplifies for the Lucchin--Matarrese attractor, to which we shall refer for definiteness, since in this case
\beq
\delta \ = \ - \ 3\,\gamma^2  \ ,
\eeq
so that the contribution from the field redefinition vanishes, but in general slow-roll scenarios this last relation would not hold.

Fig.~\ref{fig:fpp} compares $f_{PP}$ to $f_{NL}^{(M)}$, within the range $\frac{\Delta}{2}<k<2\,\Delta$, for two choices of the conformal time $\eta_0$ when the expanding phase begins. The two signals are comparable in size, and the values of $\eta_0$ mostly affect the frequency of the oscillations around the original result in~\cite{maldacena}.

\begin{figure}[ht]
\begin{center}
    \includegraphics[width=3.6in]{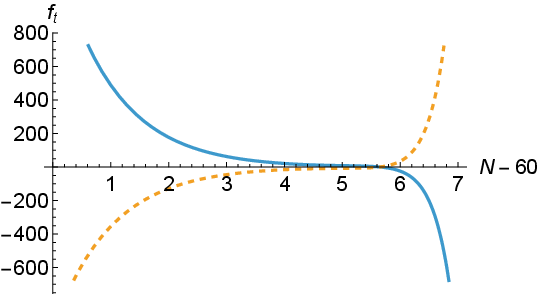}
\end{center}
\caption{The typical behavior of $f_t$ for $k=1.2\,\Delta$ (blue, solid) and for $k=2\,\Delta$ (orange, dashed) and $\epsilon=0.03$, as the number of $e$-folds increases from 60 to 67. Note the plateau present in the region $63<N<65$. This is due to the combined effect of $\langle O_{1}\rangle$ and $\langle O_{3}\rangle$, which decrease in absolute value for increasing values of $N$, and to the opposite behavior of $\langle O_4 \rangle$.}
\label{fig:fx}
\end{figure}

Figs.~\ref{fig:ft1}, ~\ref{fig:ft2} and~\ref{fig:ft3} compare the behavior of $f_t$, within the range $\frac{\Delta}{2}<k<2\,\Delta$, for different numbers of $e$--folds in the interval $61 \leq N \leq 66$. The solid curves refer to $\eta_t= -\,\frac{2}{\Delta}$ and the dashed ones to $\eta_t= -\,\frac{4}{\Delta}$. Note that for $63\leq N \leq 65$ the values of $f_t$ are appreciable but not too large and lie within ranges not excluded by the {\it Planck} Collaboration~\cite{Planck} (see also~\cite{forecasts} for forecasts related to future experiments)~\footnote{Those limits are probably too conservative for the effects under scrutiny, since they were obtained within ranges of $\ell$ that are much wider than those relevant to the present analysis. The interesting message here is the unusual emergence of upper and lower bounds on $N$ (see Fig.~\ref{fig:fx}), which could be expanded.}. Note how in this case the choice of $\eta_t$ has significant effects on the size of $f_t$. For $N<63$ or $N>65$ the peaks typically lie beyond the range identified by the {\it Planck} collaboration, so there would be a clear preference for a small window in the number of $e$-folds, if these results were to play a role in the comparison with future data. This can also be seen from fig.~\ref{fig:fx}, which shows the dependence of $f_t$ for $k= 1.2\,\Delta$ (solid line) and for $k=2\,\Delta$ (dashed line) on the number $N$ of inflationary $e$-folds, and where a plateau is clearly present for $63\leq N \leq 65$. 

In summary, $f_{PP}$ undergoes small oscillations around the original result in~\cite{maldacena}, but $f_t$ can offer some prospects for future detection and has the interesting feature of selecting a very small range of $e$-folds where our results are sizable and yet not too large.
In the present setup, as we have stressed, $\eta_0$ should be at least slightly below $\eta_t$, with both ${\cal O}\left(-\,\frac{1}{\Delta}\right)$. Let us also recall that, according to the analysis in~\cite{gkmns}, $k=\Delta$ translates into $\ell \simeq 11$ in the angular power spectrum of the CMB, which justifies the choice to concentrate our attention on the range $\frac{\Delta}{2}<k<2\,\Delta$.
All preceding results become unreliable for values of $k$ well beyond $\Delta$, since the $k \to \infty$ limit is equivalent to letting $\Delta \to 0$, which would extend the integrals to the standard range $- \infty< \eta<0$. This would require the usual contour deformation that makes it move farther and farther away from the real axis for large negative values of $\eta$, which we avoided within the limited range of interest for $k$. The modification can be simply implemented for $\eta_0$, while the way to treat the turning point is less clear to us. Note that, within the limited range for $k$ that we are focusing on, the primordial effects that we are describing can have some direct bearing on the data, without the complications due to the subsequent evolution of the Universe, as is the case for the power spectrum.

\section[Conclusions]{\sc  Conclusions}\label{sec:conclusion}
%%%%%%%%%%%%%%%%%%%%%%%%%%%%%%%%%%%%%

This paper elaborates on a setup for the three--point amplitude of curvature perturbations in scenarios inspired by high--scale supersymmetry breaking in String Theory, which involve a scalar field climbing up a steep exponential potential. 
The analysis rests on three additions to the original work in~\cite{maldacena}. The first is a scale $\Delta$~\cite{dkps,gkmns}, which introduces a low--frequency
cut in the primordial power spectrum of~\cite{cm}. If that region were accessible to us, the modification would resonate with the lack of power in the first CMB multipoles, and some evidence for this effect was collected in~\cite{gkmns}.
The scale $\Delta$ is also related to the conformal time $\eta_t$ of the turning point that completes the climbing phase, whose values are typically in the range $-\,\frac{4}{\Delta}< \eta_t< -\,\frac{1}{\Delta}$. The initial singularity, inevitably present at earlier times in the two--derivative formulation based on eq.~\eqref{2der_act}, introduces what could be termed an ultraviolet divergence in three--point amplitudes. The string corrections are supposed to overcome this problem, albeit in ways that we can only guess at this time. The tendency that manifests itself in fig.~\ref{fig:Ws} before the singular behavior takes over suggests that in a full treatment $W$ could approach the real axis from below as $\eta \to - \,\infty$. If $\epsilon$ quickly resumed a slow-roll value for $\eta<\eta_t$, the singularity would be replaced by a bounce at a conformal time $\eta_0<\eta_t$, typically also ${\cal O}\left(-\,\frac{1}{\Delta}\right)$. 

This type of scenario resonates with ideas proposed long ago in String Theory by Gasperini and Veneziano~\cite{gasp_ven}, and the Bunch--Davies condition is naturally imposed at $\eta_0$, the instant of maximum contraction for the Universe. The Schwinger--Keldysh contour of fig.~\ref{fig:contour} then yields a ``principal--part'' contribution, which undergoes small oscillations around the original result of~\cite{maldacena} at a frequency determined by $\eta_0$.
However, curvature perturbations, which appear close in spirit to the actual measurements, receive additional imprints from the turning point. These contributions can be computed in closed form relying on the ``critical'' microscopic dynamics of the climbing scalar, which dominates there, and the determination of $\Delta$ obtained in~\cite{gkmns} indicates that the resulting signal could be potentially detectable, and yet compatible with restrictions that emerged from the {\it Planck} collaboration~\cite{Planck,forecasts}, at least within the window $62<N<66$ for the number of inflationary $e$-folds. 

In analogy with the Feynman prescription for Quantum Field Theory, the $\epsilon$-prescription for the Schwinger--Keldysh contour is meant to switch off all contributions in the far past. Here we start from a finite value $\eta_0$ of the conformal time, which is typically a number of order one in units of $-\,\frac{1}{\Delta}$, and consequently the amplitudes thus obtained are only reliable for momenta that are comparable to $\Delta$, but not far beyond. In terms of CMB multipoles, $k \simeq \Delta$ corresponds to $\ell \simeq 11$~\cite{gkmns}, so the results should not be taken at face value much beyond $\ell=20$. 

As we have stressed, the modifications that we were led to introduce at very early times yield well--defined results but lack a compelling dynamical justification. They also have the virtue of leading to a picture that lies in slow--roll regime. Within their limitations, the present results add some novel features to the scenarios explored for the three--point amplitude of curvature perturbations, relying effectively on a single parameter $\Delta$ that does not appear foreign to the CMB~\cite{gkmns}.

The analysis can be extended to amplitudes involving tensor perturbations. These receive no contributions from the turning point and therefore undergo small oscillations around the original results in~\cite{maldacena}, like the ``principal part'' contribution $f_{PP}$ to the three--point function of curvature perturbations of fig.~\ref{fig:fpp}.

%%%%%%%%%%%%%%%%%%%%%%%%%%%%%%
\section*{\sc Acknowledgments}
%%%%%%%%%%%%%%%%%%%%%%%%%%%%%%
\vskip 12pt
We are grateful to N.~Bartolo, P.~Natoli, M.~Peloso, and especially to E.~Dudas, A.~Gruppuso, S.~Patil and G.L.~Pimentel, for stimulating discussions. This work was supported in part by Scuola Normale and in part by INFN (IS GSS-Pi).


\newpage
\begin{thebibliography}{99}

\bibitem{so1616} 
L.~J.~Dixon and J.~A.~Harvey,
%``String Theories in Ten-Dimensions Without Space-Time Supersymmetry,''
Nucl. Phys. B \textbf{274} (1986), 93.
%doi:10.1016/0550-3213(86)90619-X
%390 citations counted in INSPIRE as of 01 Jun 2021
L.~Alvarez-Gaume, P.~H.~Ginsparg, G.~W.~Moore and C.~Vafa,
%``An O(16) x O(16) Heterotic String,''
Phys. Lett. B \textbf{171} (1986), 155.
%doi:10.1016/0370-2693(86)91524-8
%313 citations counted in INSPIRE as of 01 Jun 2021

\bibitem{susy95}
A.~Sagnotti,
%``Some properties of open string theories,''
[arXiv:hep-th/9509080 [hep-th]];
%179 citations counted in INSPIRE as of 31 May 2021
A.~Sagnotti,
%``Surprises in open string perturbation theory,''
Nucl. Phys. B Proc. Suppl. \textbf{56} (1997), 332
%doi:10.1016/S0920-5632(97)00344-7
[arXiv:hep-th/9702093 [hep-th]].
%157 citations counted in INSPIRE as of 31 May 2021

\bibitem{usp32}
S.~Sugimoto,
%``Anomaly cancellations in type I D9-D9-bar system and the USp(32)  string
%theory,''
Prog.\ Theor.\ Phys.\  {\bf 102} (1999) 685 [arXiv:hep-th/9905159].
%%CITATION = HEP-TH 9905159;%%

\bibitem{stringtheory}
M.~B.~Green, J.~H.~Schwarz and E.~Witten, ``Superstring Theory'', 2 vols., Cambridge Univ. Press (1987); \\
J.~Polchinski, ``String theory'', 2 vols. Cambridge, UK: Cambridge Univ. Press (1998);  \\
C.~V.~Johnson, ``D-branes,'' Cambridge Univ. Press (2003); \\
B.~Zwiebach, ``A first course in string theory,'' Cambridge Univ. Press (2004); \\
K.~Becker, M.~Becker and J.~H.~Schwarz,
``String theory and M-theory: A modern introduction'' Cambridge, UK: Cambridge Univ.
Press (2007); \\
E.~Kiritsis, ``String theory in a nutshell,'' Princeton Univ. Press (2007);\\
P.~West, ``Introduction to strings and branes,'' Cambridge Univ. Press (2012).
  %%CITATION = INSPIRE-1190041;%%

\bibitem{orientifolds}
A.~Sagnotti, 
%``Open Strings And Their Symmetry Groups,'' 
in Cargese '87, ``Non-Perturbative Quantum Field
Theory'', eds. G. Mack et al (Pergamon Press, 1988), p. 521,
arXiv:hep-th/0208020;
%%CITATION = HEP-TH 0208020;%%
G.~Pradisi and A.~Sagnotti,
%``Open String Orbifolds,''
Phys.\ Lett.\ {\bf B 216} (1989) 59;
%%CITATION = PHLTA,B216,59;%%
P.~Horava,
%``Strings On World Sheet Orbifolds,''
Nucl.\ Phys.\ {\bf B 327} (1989) 461;
%%CITATION = NUPHA,B327,461;%%
P.~Horava, 
%``Background Duality Of Open String Models,''
Phys.\ Lett.\ {\bf B 231} (1989) 251;
%%CITATION = PHLTA,B231,251;%%
M.~Bianchi and A.~Sagnotti,
%``On The Systematics Of Open String Theories,''
Phys.\ Lett.\ {\bf B 247} (1990) 517;
%%CITATION = PHLTA,B247,517;%%
M.~Bianchi and A.~Sagnotti,
%``Twist Symmetry And Open String Wilson Lines,''
Nucl.\ Phys.\ {\bf B 361} (1991) 519;
%%CITATION = NUPHA,B361,519;%%
M.~Bianchi, G.~Pradisi and A.~Sagnotti,
%``Toroidal compactification and symmetry breaking in open string theories,''
Nucl.\ Phys.\ {\bf B 376} (1992) 365;
%%CITATION = NUPHA,B376,365;%%
A.~Sagnotti,
 %``A Note on the Green-Schwarz mechanism in open string theories,''
 Phys.\ Lett.\  {\bf B 294} (1992) 196
 [arXiv:hep-th/9210127].\\
 %%CITATION = PHLTA,B294,196;%%
 For reviews, see:
E.~Dudas,
%``Theory and phenomenology of type I strings and M-theory,''
Class.\ Quant.\ Grav.\  {\bf 17} (2000) R41 [arXiv:hep-ph/0006190];
%%CITATION = HEP-PH 0006190;%%
C.~Angelantonj and A.~Sagnotti,
%``Open strings,''
Phys.\ Rept.\  {\bf 371} (2002) 1 [Erratum-ibid.\  {\bf 376} (2003)
339] [arXiv:hep-th/0204089];
%%CITATION = HEP-TH 0204089;%%
C.~Angelantonj and I.~Florakis,
%``A Lightning Introduction to String Theory,''
%doi:10.1007/978-981-19-3079-9\_53-1
[arXiv:2406.09508 [hep-th]].

\bibitem{climbing}
E.~Dudas, N.~Kitazawa and A.~Sagnotti,
%``On Climbing Scalars in String Theory,''
Phys. Lett. B \textbf{694} (2011), 80
%doi:10.1016/j.physletb.2010.09.040
[arXiv:1009.0874 [hep-th]].

\bibitem{dm}
E.~Dudas and J.~Mourad,
% ``Brane solutions in strings with broken supersymmetry and dilaton tadpoles,''
 Phys.\ Lett.\  {\bf B 486} (2000) 172
 [arXiv:hep-th/0004165].
 %%CITATION = PHLTA,B486,172;%%

\bibitem{russo}
J.~G.~Russo,
 % ``Exact solution of scalar-tensor cosmology with exponential potentials and transient acceleration,''
  Phys.\ Lett.\  {\bf B 600} (2004), 185
  [arXiv:hep-th/0403010].
  %%CITATION = PHLTA,B600,185;%%

  \bibitem{starobinsky}
   A.~A.~Starobinsky,
 % ``A New Type of Isotropic Cosmological Models Without Singularity,''
  Phys.\ Lett.\ {\bf B 91} (1980) 99.
 %%CITATION = PHLTA,B91,99;%%

 \bibitem{inflation}
 A.~A.~Starobinsky,
%  ``A New Type of Isotropic Cosmological Models Without Singularity,''
  Phys.\ Lett.\ {\bf B 91} (1980) 99;
 %%CITATION = PHLTA,B91,99;%%
  D.~Kazanas,
 %``Dynamics of the Universe and Spontaneous Symmetry Breaking,''
  Astrophys.\ J.\  {\bf 241} (1980) L59;
 %%CITATION = ASJOA,241,L59;%%
  K.~Sato,
%``Cosmological Baryon Number Domain Structure and the First Order Phase Transition of a Vacuum,''
  Phys.\ Lett.\  {\bf B 99} (1981) 66;
 %%CITATION = PHLTA,B99,66;%%
 A.~H.~Guth,
% ``The Inflationary Universe: A Possible Solution to the Horizon and Flatness Problems,''
 Phys.\ Rev.\ {\bf D 23} (1981) 347;
  %%CITATION = PHRVA,D23,347;%%
 A.~D.~Linde,
% ``A New Inflationary Universe Scenario: A Possible Solution of the Horizon, Flatness, Homogeneity, Isotropy and Primordial Monopole Problems,''
  Phys.\ Lett.\ {\bf B 108} (1982) 389;
  %%CITATION = PHLTA,B108,389;%%
   A.~Albrecht and P.~J.~Steinhardt,
%``Cosmology for Grand Unified Theories with Radiatively Induced Symmetry Breaking,''
  Phys.\ Rev.\ Lett.\  {\bf 48} (1982) 1220;
   A.~D.~Linde,
%``Chaotic Inflation,''
  Phys.\ Lett.\ {\bf B 129} (1983) 177.
 %%CITATION = PHLTA,B129,177;%%
For reviews, see: 
 N.~Bartolo, E.~Komatsu, S.~Matarrese and A.~Riotto,
%``Non-Gaussianity from inflation: Theory and observations,''
  Phys.\ Rept.\  {\bf 402} (2004) 103
  [astro-ph/0406398].\\
  %%CITATION = ASTRO-PH/0406398;%%
V.~Mukhanov,
  ``Physical foundations of cosmology,''
  Cambridge Univ. Press (2005); \\
S.~Weinberg, 
``Cosmology,''
 Oxford Univ. Press (2008); \\
D.~H.~Lyth and A.~R.~Liddle,
``The primordial density perturbation: Cosmology, inflation and the origin of structure,''
  Cambridge Univ. Press (2009); \\
  D.~S.~Gorbunov and V.~A.~Rubakov,
  ``Introduction to the theory of the early universe: Cosmological perturbations and inflationary theory,''
  World Scientific (2011);\\
  %doi:10.1142/7874
  %%CITATION = doi:10.1142/7874;%%
  %15 citations counted in INSPIRE as of 04 Aug 2017
 J.~Martin, C.~Ringeval and V.~Vennin,
 % ``Encyclopaedia Inflationaris,''
  Phys.\ Dark Univ.\  {\bf 5-6} (2014) 75
  [arXiv:1303.3787 [astro-ph.CO]];
  %%CITATION = doi:10.1016/j.dark.2014.01.003;%%
    N.~Vittorio, ``Cosmology,''
 CRC Press (2018);
  D.~Baumann, ``Cosmology,''
 Cambridge Univ. Press (2022). \\

\bibitem{dkps}
E.~Dudas, N.~Kitazawa, S.~P.~Patil and A.~Sagnotti,
%``CMB Imprints of a Pre-Inflationary Climbing Phase,''
JCAP \textbf{05} (2012), 012
%doi:10.1088/1475-7516/2012/05/012
[arXiv:1202.6630 [hep-th]];
N.~Kitazawa and A.~Sagnotti,
%``Pre-inflationary clues from String Theory?,''
JCAP \textbf{04} (2014), 017
%doi:10.1088/1475-7516/2014/04/017
[arXiv:1402.1418 [hep-th]].

\bibitem{cm}
V.~F.~Mukhanov and G.~V.~Chibisov,
% ``Quantum Fluctuations and a Nonsingular Universe,''
JETP Lett. \textbf{33} (1981), 532.
%1682 citations counted in INSPIRE as of 03 Jun 2021
For a review, see:
V.~F.~Mukhanov, H.~A.~Feldman and R.~H.~Brandenberger,
%``Theory of cosmological perturbations. Part 1. Classical perturbations. Part 2. Quantum theory of perturbations. Part 3. Extensions,''
Phys. Rept. \textbf{215} (1992), 203.
%doi:10.1016/0370-1573(92)90044-Z
%3462 citations counted in INSPIRE as of 27 Nov 2023

\bibitem{peloso}
C.~R.~Contaldi, M.~Peloso, L.~Kofman and A.~D.~Linde,
%``Suppressing the lower multipoles in the CMB anisotropies,''
JCAP \textbf{07} (2003), 002
%doi:10.1088/1475-7516/2003/07/002
[arXiv:astro-ph/0303636 [astro-ph]].
%395 citations counted in INSPIRE as of 19 Sep 2025

\bibitem{gkmns}
A.~Gruppuso and A.~Sagnotti,
%``Observational Hints of a Pre--Inflationary Scale?,''
Int. J. Mod. Phys. D \textbf{24} (2015) no.12, 1544008
%doi:10.1142/S0218271815440083
[arXiv:1506.08093 [astro-ph.CO]];
%31 citations counted in INSPIRE as of 11 Sep 2025
A.~Gruppuso, N.~Kitazawa, N.~Mandolesi, P.~Natoli and A.~Sagnotti,
%``Pre-Inflationary Relics in the CMB?,''
Phys. Dark Univ. \textbf{11} (2016), 68
%doi:10.1016/j.dark.2015.12.001
[arXiv:1508.00411 [astro-ph.CO]];
A.~Gruppuso, N.~Kitazawa, M.~Lattanzi, N.~Mandolesi, P.~Natoli and A.~Sagnotti,
%``The Evens and Odds of CMB Anomalies,''
Phys. Dark Univ. \textbf{20} (2018), 49
%doi:10.1016/j.dark.2018.03.002
[arXiv:1712.03288 [astro-ph.CO]].

\bibitem{review_pst}
For a recent review, see: D.~Baumann, D.~Green, A.~Joyce, E.~Pajer, G.~L.~Pimentel, C.~Sleight and M.~Taronna,
%``Snowmass White Paper: The Cosmological Bootstrap,''
SciPost Phys. Comm. Rep. \textbf{2024} (2024), 1
%doi:10.21468/SciPostPhysCommRep.1
[arXiv:2203.08121 [hep-th]].
%170 citations counted in INSPIRE as of 25 Sep 2025

\bibitem{gasp_ven}
M.~Gasperini and G.~Veneziano,
%``Pre - big bang in string cosmology,''
Astropart. Phys. \textbf{1} (1993), 317-339
%doi:10.1016/0927-6505(93)90017-8
[arXiv:hep-th/9211021 [hep-th]];
%1017 citations counted in INSPIRE as of 09 Sep 2025
M.~Gasperini and G.~Veneziano,
%``The Pre - big bang scenario in string cosmology,''
Phys. Rept. \textbf{373} (2003), 1-212
%doi:10.1016/S0370-1573(02)00389-7
[arXiv:hep-th/0207130 [hep-th]].
%818 citations counted in INSPIRE as of 09 Sep 2025

\bibitem{maldacena}
J.~M.~Maldacena,
%``Non-Gaussian features of primordial fluctuations in single field inflationary models,''
JHEP \textbf{05} (2003), 013
%doi:10.1088/1126-6708/2003/05/013
[arXiv:astro-ph/0210603 [astro-ph]].
%3023 citations counted in INSPIRE as of 06 Jul 2025

\bibitem{reviews}
J.~Noller and J.~Magueijo,
%``Non-Gaussianity in single field models without slow-roll,''
Phys. Rev. D \textbf{83} (2011), 103511
%doi:10.1103/PhysRevD.83.103511
[arXiv:1102.0275 [astro-ph.CO]];
%40 citations counted in INSPIRE as of 06 Jul 2025
H.~Collins,
%``Primordial non-Gaussianities from inflation,''
[arXiv:1101.1308 [astro-ph.CO]].
%35 citations counted in INSPIRE as of 06 Jul 2025

\bibitem{Planck}
P.A.R.~Ade \textit{et al.} [Planck],
%``Planck 2013 Results. XXIV. Constraints on primordial non-Gaussianity,''
Astron. Astrophys. \textbf{571} (2014), A24
%doi:10.1051/0004-6361/201321554
[arXiv:1303.5084 [astro-ph.CO]];
%840 citations counted in INSPIRE as of 15 Sep 2025
P.~A.~R.~Ade \textit{et al.} [Planck],
%``Planck 2015 results. XVII. Constraints on primordial non-Gaussianity,''
Astron. Astrophys. \textbf{594} (2016), A17
%doi:10.1051/0004-6361/201525836
[arXiv:1502.01592 [astro-ph.CO]].
%862 citations counted in INSPIRE as of 15 Sep 2025
Y.~Akrami \textit{et al.} [Planck],
%``Planck 2018 results. IX. Constraints on primordial non-Gaussianity,''
Astron. Astrophys. \textbf{641} (2020), A9
%doi:10.1051/0004-6361/201935891
[arXiv:1905.05697 [astro-ph.CO]].
%904 citations counted in INSPIRE as of 15 Sep 2025

\bibitem{forecasts}
M.~Abitbol \textit{et al.} [Simons Observatory],
%``The Simons Observatory: science goals and forecasts for the enhanced Large Aperture Telescope,''
JCAP \textbf{08} (2025), 034
%doi:10.1088/1475-7516/2025/08/034
[arXiv:2503.00636 [astro-ph.IM]];
%38 citations counted in INSPIRE as of 15 Sep 2025
W.~Sohn and J.~Fergusson,
%``CMB-S4 forecast on the primordial non-Gaussianity parameter of feature models,''
Phys. Rev. D \textbf{100} (2019) no.6, 063536
%doi:10.1103/PhysRevD.100.063536
[arXiv:1902.01142 [astro-ph.CO]].
%15 citations counted in INSPIRE as of 15 Sep 2025

\bibitem{fss}
P.~Fr\'e, A.~Sagnotti and A.~S.~Sorin,
%``Integrable Scalar Cosmologies I. Foundations and links with String Theory,''
Nucl. Phys. B \textbf{877} (2013), 1028
%doi:10.1016/j.nuclphysb.2013.10.015
[arXiv:1307.1910 [hep-th]];
A.~Sagnotti,
%``Brane SUSY breaking and inflation: implications for scalar fields and CMB distortion,''
Phys. Part. Nucl. Lett. \textbf{11} (2014), 836
%doi:10.1134/S1547477114070395
[arXiv:1303.6685 [hep-th]].
%34 citations counted in INSPIRE as of 27 Jun 2025

\bibitem{lm}
 F.~Lucchin and S.~Matarrese,
%``Power Law Inflation,''
Phys. Rev. D \textbf{32} (1985), 1316.
% doi:10.1103/PhysRevD.32.1316
%944 citations counted in INSPIRE as of 15 Feb 2024

\bibitem{sub}
J.~Martin, H.~Motohashi and T.~Suyama,
%``Ultra Slow-Roll Inflation and the non-Gaussianity Consistency Relation,''
Phys. Rev. D \textbf{87} (2013) no.2, 023514
%doi:10.1103/PhysRevD.87.023514
[arXiv:1211.0083 [astro-ph.CO]];
%350 citations counted in INSPIRE as of 01 Oct 2025
A.~Negro and S.~P.~Patil,
%``An {\'E}tude on the regularization and renormalization of divergences in primordial observables,''
Riv. Nuovo Cim. \textbf{47} (2024) no.3, 179-228
%doi:10.1007/s40766-024-00053-0
[arXiv:2402.10008 [hep-th]];
%9 citations counted in INSPIRE as of 01 Oct 2025
J.~Chluba, J.~Hamann and S.~P.~Patil,
%``Features and New Physical Scales in Primordial Observables: Theory and Observation,''
Int. J. Mod. Phys. D \textbf{24} (2015) no.10, 1530023
%doi:10.1142/S0218271815300232
[arXiv:1505.01834 [astro-ph.CO]].
%253 citations counted in INSPIRE as of 01 Oct 2025
\end{thebibliography}
\end{document}